\documentclass[final,5p,times,twocolumn]{elsarticle}
\usepackage{graphics}
\usepackage{amssymb}
\biboptions{compress}

\journal{Nuclear Instruments and Methods A}

\begin{document}

\begin{frontmatter}


%%\ead{stuhl@atomki.hu}

\title{A pair spectrometer for measuring multipolarities of energetic 
nuclear transitions}
%%\author[label1,label2]{<author name>}
%%\address[label1]{<address>}

\author[label1]{J.~Guly\'as}
\address[label1]{MTA-ATOMKI, Institute for Nuclear Research, Hungarian Academy 
of Sciences}
\address[label2]{Nikhef National Institute for Subatomic Physics,
Science Park 105, 1098 XG Amsterdam, The Netherlands}
\author[label2]{T.J.~Ketel}
\author[label1]{A.J.~Krasznahorkay}
\ead{kraszna@atomki.hu}

\author[label1]{M.~Csatl\'os}
\author[label1]{L.~Csige}
\author[label1]{Z.~G\'acsi}
\author[label1]{M.~Hunyadi}
\address[label3]{CERN,European Organization for Nuclear Research, 
Geneva, Switzerland}
\author[label3]{A.~Krasznahorkay}
\author[label1]{A.~Vit\'ez}
\author[label1]{T.G.~Tornyi}

\begin{abstract}

\indent 

A multi-detector array has been designed and constructed for the simultaneous
measurement of energy- and angular correlations of electron-positron
pairs. Experimental results are obtained over a wide angular range for
high-energy transitions in $^{16}$O, $^{12}$C and $^8$Be. A comparison with
GEANT simulations demonstrates that angular correlations between 50 and 180
degrees of the e$^+$e$^−$ pairs in the energy range between 6 and 18 MeV can
be determined with sufficient resolution and efficiency. \end{abstract}
%At the same time, a
%significant anomaly is observed in the angular correlation measured for the
%18.1 MeV $M1$ transition in $^8$Be.


\end{frontmatter}

%\begin{linenumbers}

\section{Introduction}

Spectroscopy of internal pair conversion (IPC) has a long tradition
\cite{wi65,ro50}. In a wide range of energies and atomic numbers, the
conversion coefficient for internal electron-positron pair formation are
fairly high, typically in the order of $10^{-4}$-$10^{-3}$ \cite{sc79}. The
measurement of these coefficients offers an effective method for determining
the multipolarity of electromagnetic transitions (especially of high-energy
and low-multipolarity transitions) \cite{sc81}.

The determination of the multipolarity of the high-energy transitions produced
after particle capture reactions might be especially important for nuclear
astrophysics to gain deeper understanding of the dynamics of capture processes
leading to a more accurate and reliable extraction of the astrophysical $S$
factor and the thermonuclear reactivity \cite{ch94,sp99}.

In many light nuclei, the cross section for the radiative capture of protons,
neutrons, deuterons and $\alpha$ particles has been observed to consist of a
background slowly varying with beam energy, upon which the various known
resonances in the reaction are superposed. This smooth background, which is
important for nuclear astrophysics, has been identified as an extra-nuclear
channel phenomenon, since the process takes place far from the nucleus (40-50
fm) rather than in the nuclear interior \cite{ro73}. This has been designated
as direct capture (single-step) reaction. The direct capture process
represents a transition for the projectile from an initial continuum state to
a final state via interaction with the electromagnetic field. Usually, it has
a strong $E1$ component but other multipolarities can also contribute. For the
extraction and extrapolation of the astrophysical $S$ factor, it is important
to know the multipole composition of such background radiations
\cite{ch94,sp99}.

The multidetector array is also designed to search for deviations from IPC due
to the creation and subsequent decay into electron-positron pairs of a
hypothetical short-lived neutral boson. Recent results from both underground
and cosmic ray experiments suggest that dark matter may be explained by a
light boson having a mass of 10 MeV - 10 GeV and coupled to electrons and
positrons. There have been several attempts to observe evidence for such
particles, using data from running facilities
\cite{me11,ab11,le12,ec12,ar12,ba13,ad13,ba08} or re-analyzing old experiments
\cite{bj09,an12,bl11,gn12,gni12}. Since no evidence for their existence was
found, limits on its coupling to ordinary matter were set as a function of its
mass. In the near future, new experiments are expected to extend those limits
in a region of couplings and/or masses so far unexplored. It is not widely
known, but indications were found for the existence of such a light boson also
in some nuclear physics experiments. While anomaly was observed in the
internal pair production, the overall results were not consistent with the
involvement of a neutral boson \cite{bo96,bo97,bo01}. A limit of
$\leq4.1\times$10$^{-4}$ was obtained for the boson to $\gamma$-ray branching
ratio \cite{bo96,bo97,sa88,bo01,ti04}.

\section{Internal Pair Creation (IPC)}

\begin{figure}[ht]
\centering \includegraphics[width=80mm]{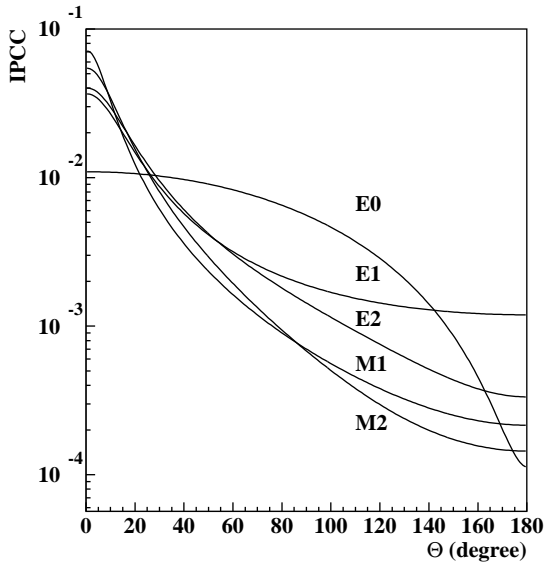}
\caption{\label{fig:ang-rose}Calculated angular correlations of e$^+$e$^-$
  pairs obtained from IPC for different multipolarities and a transition
  energy of E$_\gamma$=17 MeV.}
\end{figure}

Quantum electrodynamics (QED) predicts \cite{wi65,ro50} that the angular
correlation between the e$^+$e$^-$ pairs (emitted in IPC) peaks at 0$^{\circ}$
and drops rapidly with the correlation angle ($\Theta$) as shown in
Fig.~\ref{fig:ang-rose}.

The above calculations show that the angular correlations at small separation
angles are almost independent of the multipolarity of the radiation, whereas
at large separation angles, they depend critically upon the multipole
order. Thus, it is important to measure angular correlations efficiently at
large angles.

\section{The two-body decay of a boson}

When a nuclear transition occurs by emission of a short-lived ($\tau$ $<$
10$^{-13}$ s) neutral particle, the annihilation into an e$^+$e$^-$ pair is
anti-parallel (i.e. $\Theta_{cm}=$180$^{\circ}$) in the center of mass
system. In the laboratory system, their angular distribution is peaked
($\Delta\Theta<10^{\circ}$) at intermediate angles due to the Lorentz boost
and provides an unique signature for the existence and a measure for the mass
of an intermediate boson. In order to search for such an anomaly in the
angular correlation, we need a spectrometer with sufficient angular
resolution.

The invariant mass can be determined approximately from the relative angle
$\Theta$ between e$^+$ and e$^-$ and from their energies in the following
way\cite{bo97}:

\begin{equation}
 m^2\approx(1-y^2)E^2\sin^2(\Theta/2),
\end{equation}

\noindent where $E=E^++E^-+1.022$ MeV is the transition energy and
$y=(E^+-E^-)/(E^++E^-)$, with $E^{+(-)}$ indicating the kinetic energy of the
positron (electron) in the laboratory system.

\section{Overview of pair spectrometers}

Magnetic $\beta$ ray spectrometers were used first for internal pair formation
studies \cite{da54,be55,al58,wa64,kj58}. Maximal detection efficiency of
10$^{-4}$ for electron-positron pair detection was achieved for a few cases
\cite{al58,kj58}. Improvement of the pair resolution by improvement of the
momentum resolution (to 1.3\%) with smaller particle transmission reduced the
efficiency to 5x10$^{-6}$. An important advance \cite{wa64} in the use of
intermediate-image pair spectrometer was provided by the installation of a
specially designed spiral baffle system which selected electron-positron
internal pairs emitted at large relative angles
($50^\circ\leq\theta\leq90^\circ$).

The next generation of internal-pair spectrometers used two $dE/dx+E$
scintillator-detector telescopes for the detection of the electron-positron
pairs in quadruple coincidence \cite{ad74,ul77}. A multi-detector (six
scintillation electron telescopes plus an annular Si(Li) particle detector)
high-efficiency pair spectrometer was built by Birk and co-workers
\cite{bi82}. An experimental pair-line efficiency of 28\% and a sum-peak
energy resolution of 12\% for the 6.05 MeV E0 pair line in $^{16}$O were
achieved.

Schumann and Waldschmidt have detected internal pair spectra in the energy
range of 2.8-6.5 MeV from an (n,$\gamma$) reaction with a combination
super-conducting solenoid transporter plus Si(Li)-detector spectrometer
\cite{sc74}. The pair-line efficiency of the spectrometer \cite{wa70} was
large, but it had a very limited discrimination power for different
multipolarities in this energy region.

The Debrecen superconducting solenoid transporter plus two-Si(Li)-de\-tec\-tor
electron spectrometer was also adapted for internal-pair studies
\cite{pa84}. The observed pair-line efficiency for two detectors operated in
sum-coincidence mode was 35\%, while the energy resolution was 0.6\% at 2
MeV. A similar spectrometer built by Kib\'edi and co-workers \cite{ki90} and
has been used recently for internal pair studies \cite{ki12}.

A highly segmented phoswich array of plastic scintillators was constructed for
measurements of e$^+$e$^-$ pairs emitted in high-energy electromagnetic
transitions in nuclei by Montoya and co-workers \cite{mo93}. Electron
(positron) energies of 2-30 MeV can be measured by each individual element,
with a total transition energy resolution of $\delta$E/E = 13\% for a 20 MeV
transition. The array covers 29\% of the full solid angle and its efficiency
is 1.6\% for a 6 MeV $E0$ internal pair decay, and 1.1\% for an 18 MeV $E1$
transition.

A positron-electron pair spectroscopy instrument (PEPSI) was designed to
measure transitions in the energy region of 10-40 MeV by Buda and co-workers
\cite{bu93}. It consists of Nd$_2$Fe$_{14}$B permanent magnets forming a
compact 4$\pi$ magnetic filter consisting of 12 positron and 20 electron
mini-orange-like spectrometers.

A $\Delta E-E$ multi-detector array was constructed by Stiebing and co-workers
\cite{st04} from plastic scintillators for the simultaneous measurement of
energy and angular correlation of e$^+$e$^-$ pairs produced in internal pair
conversion (IPC) of nuclear transitions up to 18 MeV. The array was designed
to search for deviations from IPC stemming from the creation and subsequent
decay into e$^+$e$^-$ pairs of a hypothetical short-lived neutral boson. The
angular resolution of the spectrometer determined by the solid angle of the
telescopes was $\Delta\Theta=15^\circ$, while the efficiency for one pair of
telescopes: $\approx3\times 10^{-5}$. The investigated angular range was
extended from 20$^\circ$ to 131$^\circ$.

In this paper, we present a novel e$^+$e$^{-}$ pair spectrometer equipped with
multi-wire proportional chambers and large volume plastic scintillator
telescopes having remarkably higher efficiency and better angular resolution
than previously obtained by Stiebing and co-workers \cite{st04}.

\section{Monte-Carlo simulations}

Monte Carlo (MC) simulations of the experiment were performed using the GEANT3
code in order to determine the detector response function. For different
transition energy and multipolarity a lookup table is created for electron an
positron energies and correlation angle using the Rose calculations
\cite{ro50}. 
 The first electrons (or positrons) are generated isotropically, 
with $\phi_e$ random between 0 and 2$\pi$
and $\theta_e$ as a sine distribution, and 
the second particles with relative angles $\phi$ and $\theta$, with $\theta$
according to the lookup table.  Isotropic emission of pairs would also result
in a sine distribution for the relative angles $\theta$, the so-called
correlation angle.

Also boson decays can be generated as well as gamma ray coincidences. The
electrons and positrons are followed through the setup and the detected energy
losses are stored, including detection of annihilation radiation from the
stopped positrons. The energy loss steps are small until a final energy of 90
keV. The simulated events are stored in a similar way as the measurements, but
now as precise deposited energies and positions inside the wire chambers and
including the generated electron and positron energies and correlation angles.

\section{The spectrometer}

\begin{figure}[ht]
\centering
\includegraphics[width=70mm]{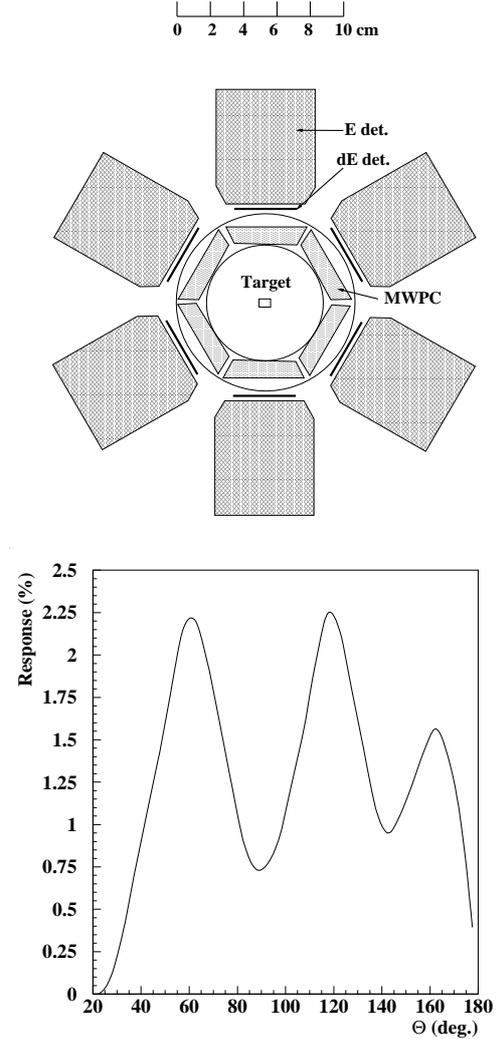}
\caption{\label{fig:telescope6m} (top) Initial arrangement with six telescopes
and (bottom) detection response as a function of the correlation angle
between the e$^+$e$^-$ pairs in Monte Carlo simulations.}
\end{figure}

Plastic scintillator detectors combine reasonable energy resolution with
minimum response to $\gamma$ radiation and with excellent characteristics for
fast, sub-nanosecond coincidence timing, which is crucial for good background
reduction. Thus, we use plastic $\Delta$E-E detector telescopes for the
detection of the e$^+$e$^-$. In contrast to Ref.~\cite{st04}, very thin
$\Delta E$ detectors (52$\times$52$\times$1 mm$^3$) were chosen that gives a
remarkably improved $\gamma$ suppression. The $E$ detectors have similar
dimensions (80$\times$60$\times$70 mm$^3$) as in Ref.~\cite{st04}. The
spectrometer setup is shown in Fig.~\ref{fig:telescope6m} with six
scintillation detector telescopes and six position sensitive gaseous detectors
at 60 degrees relative to their neighbors surrounding the target inside the
carbon fiber beam pipe.  The response of the detector set-up as a function of
correlation angle theta for isotropic emission of e+e- pairs is shown in
Fig. 2 (bottom).  A detector with 4$\pi$ solid angle acceptance would show a
sine distribution and the simulated curve with three sharp peaks can be
understood as the limited phase space with only detector combinations at 60,
120, and 180 degrees with an angular range in a single detector of about 40
degrees.  Another setup with five telescopes will be also described with a
smoother acceptance for the angular correlation of the e$^+$e$^-$ pairs.

$\gamma$ rays were detected by a Ge clover detector at a distance of 25 cm
from the target behind the Faraday-cup. The detector has an active volume of
470 cm$^3$ and it is also equipped with a BGO anti-coincidence shield
\cite{el03}.

The positions of the hits are measured by multiwire proportional counters
(MWPC) which was constructed at ATOMKI based on the concept of
Ref.~\cite{ch79} and placed in front of the $\Delta E$ and $E$ detectors. The
anode of the MWPC is a set of parallel 10 $\mu$m thick gold-plated tungsten
wires at a distance of 2 mm from each other. The two cathodes are composed of
silver-plated copper wires having a diameter of 0.1 mm and separated by 1.27
mm. The anode-cathode distance is 3.5 mm. The two cathodes are placed
perpendicularly to each other giving the $x$ and $y$ coordinates of the
hit. Delay-line read-out (10 ns/taps) is used for the cathode
wires. Ar(80\%)+CO$_2$(20\%) counting gas was flowing across the detector
volume at atmospheric pressure. The accuracy of the $(x,y)$ coordinates
implies an angular resolution of $\Delta\Theta = 2^\circ$ (FWHM) in the
40$^\circ$-180$^\circ$ range, which is approximately five times better than in
Ref.~\cite{st04}. The efficiency of the MWPC detectors was estimated to be
80\%.

\subsection{Beam and Target}

To minimize the amount of material around the target, a 24 cm long
electrically conducting carbon fiber tube with a radius of 3.5 cm and a wall
thickness of 0.8 mm is used. The target, positioned perpendicular to the beam,
is mounted on a target holder supported from the back by two perspex rods of 3
mm diameter. The original 0.5 mm thick Al target holders with 10 mm inner
diameter opening were replaced when data showed shadowing due to scattering in
the aluminum sides.The GEANT simulations confirmed this shadowing and also
some background via external pair production. To avoid this, the targets were
evaporated onto 10-micron thick, 50-mm long and 5 mm wide Al strips, which was
stretched between two 3-mm thick Plexiglas rods. The rods are arranged
parallel to the beam and their distance from the beam was 25 mm. On the basis
of simulations, they did not cause significant background via external
electron-positron conversion. The bars were 12-cm long, and the placement was
done so as not to cause any shadowing effects in any of the telescopes. The
targets have a typical thickness of 0.3 mg/cm$^2$, which is adapted to the
resonance width of the reaction under investigation as well as to the demand
of a sufficient real-to-random ratio of coincidences. The beam is absorbed in
a Tantalum Faraday-cup 15 cm behind the target.

\subsection{Trigger for data readout and data-acquisition}

The signals from the photomultipliers of the $\Delta$E-E detectors are
processed in constant fraction discriminator units (CF8000). The CFD
thresholds are adjusted slightly above the noise level of the $\Delta$E
detectors (which are essentially insensitive to $\gamma$ ray events) and a bit
higher for the $E$ detectors. Chance events from double (or multiple) hits by
$\gamma$ rays in the $E$ detectors are suppressed by requiring a $\Delta$E-E
coincidence. The resulting telescope signals are analyzed by a logical unit
requiring multiplicity-2 coincidences. In order to allow the simultaneous
measurement of single telescope events, the trigger module is set to allow a
scaled-down fraction of single telescope events as well. Time and energy
signals of the $\Delta$E-E detectors as well as the time signals (Up, Down,
Left, Right) of the MWPC detector are recorded. The spectra of single
telescope events are used for on-line monitoring of the efficiencies and an
approximate energy calibration of the $E$ detectors. Especially for the
$\Delta$E detectors with their low CFD thresholds this on-line survey is
important. In the off-line analysis these spectra provide a reliable way to
determine the telescope efficiencies.

\subsection{Energy calibration of the spectrometer}

The energy calibration of the telescopes for low energies was made with the
Compton edges of a $^{60}$Co source, while at high energies we used the
Compton edges of high energy $\gamma$ transitions coming from proton capture
reactions. The high energy edges of the singles electron spectra offered also
good calibration points, which was used for on-line gain monitoring and
corrections as well. It was possible to correct the gain shifts with a
precision of about 1\%, well below the energy resolution of the
detectors. Finally, the sum energy spectra was checked for the 6.05-MeV
transition in $^{16}$O excited in the $^{19}$F(p,$\alpha$)$^{16}$O reaction,
the 4.44-MeV and 15.1-MeV transitions in $^{12}$C excited in the
$^{11}$B(p,$\gamma$)$^{12}$C reaction and the 17.6 MeV line in $^8$Be excited
in the $^{7}$Li(p,$\gamma$)$^{8}$Be reaction.

\begin{figure}[ht]
\centering
\includegraphics[width=70mm]{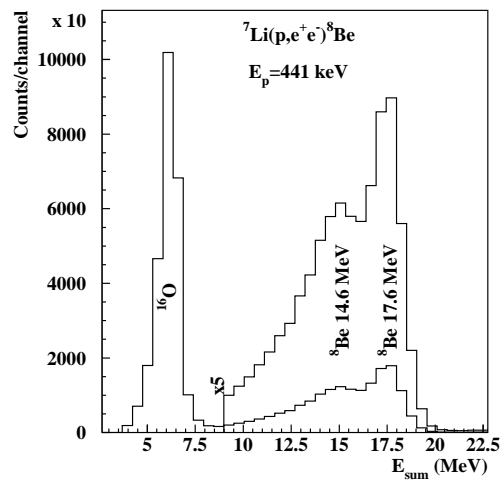}
\caption{\label{fig:441-sum}Total energy spectrum, reconstructed from the sum
  energy deposited in the scintillators, the undetected energy loss and the
  mass of the e$^-$e$^+$ pairs, produced at E$_p$=0.441 MeV using LiF$_2$
  targets.}
\end{figure}

The energy threshold settings of the $\Delta$E detectors were found to be also
very important. We had to make sure that we are not cutting too much from the
low-energy part of the energy-loss distributions. The energy calibration of
those detectors was based on the comparison of the measured and simulated
energy loss distributions of the strong 6.05 MeV $E0$ transition in
$^{16}$O. An example of such a measurement is shown in Fig.~\ref{fig:441-sum}.

\subsection{Efficiency calibration of the spectrometer}

It was crucial for the precise angular correlation measurements to measure and
understand the response to isotropic e$^+$e$^-$ pairs of the whole detector
system as a function of the correlation angle.  We were aiming at a precision
of about 1\% for the shape of the response function.

The detectors measure continuous e$^+$e$^-$ spectra and the sum of the
energies are constructed off-line. Due to the energy loss in the wall of the
chamber and in the $\Delta$E detectors, as well as the finite thresholds of
the discriminators (CFD), the low-energy part of the spectrum is always cut
out. Thresholds should be set equally to have similar efficiencies for the
different telescopes. After a proper energy calibration of the telescopes, it
was done by software cuts. The response of the MWPC detectors depends slightly
on the position of the hit, the energy of the particle and might slowly change
also in time.

\begin{figure}[ht]
\centering \includegraphics[width=70mm]{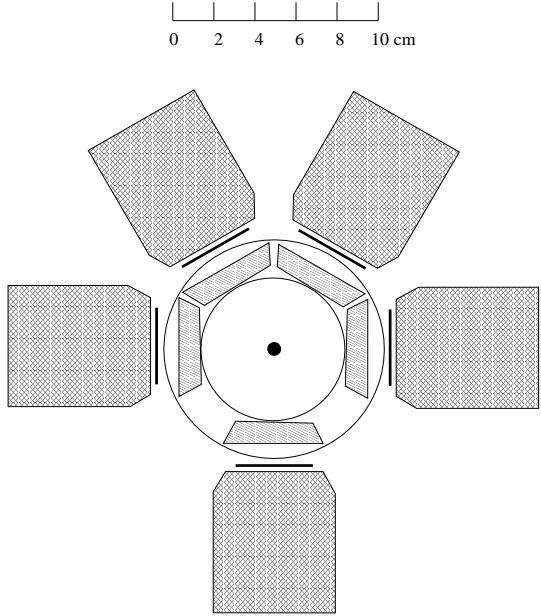}
\caption{\label{fig:telescopes5m}Final schematic arrangement with five
  telescopes to detect e$^+$e$^-$ pairs.}
\end{figure}

The response curve depends primarily on the geometrical arrangement of the
detector telescopes. As shown in Fig.~\ref{fig:telescope6m}, initially we used
six equivalent telescopes placed symmetrically around the target. However, due
to the six-fold rotational symmetry of the spectrometer and the finite solid
angle of the detectors, the response varied drastically as a function of the
correlation angle. Moreover, at the minima of the curve, the edge effects of
the detectors dominated, which made the response values under-defined. Thus,
it was advantageous to break the rotational symmetry to make the response
curve smoother. Since we also had to increase the response around 90 degrees,
we set the geometry of the setup as shown in Fig.~\ref{fig:telescopes5m}.

\begin{figure}[htb]
\centering
\includegraphics[width=70mm]{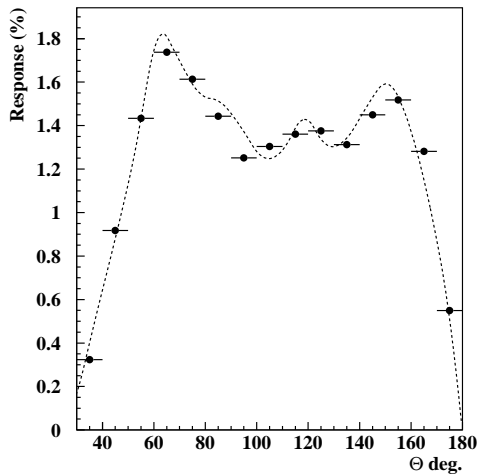}
\caption{\label{fig:effi-exp-th}
Detector response for the five-telescope setup as a function of correlation
angle ($\theta$) for isotropic emission of e$^+$e$^-$ pairs (curve) in Monte
Carlo simulations and (data points) from experimental data as explained in the
text.}
\end{figure}

Beside the e$^+$e$^-$ coincidences, the down-scaled single electron events
($\Delta E-E$ coincidences) were also collected during the whole experiment
for making experimental energy and response calibrations. An event mixing
method \cite{ra14} was used to determine experimentally the relative response
of the spectrometer as a function of the correlation angle. According to the
method, uncorrelated lepton pairs were generated from
subsequent single events and their correlation angle was calculated as for the
coincident events. The resulted angular correlation for the uncorrelated
events gave us the experimental response curve. Reasonably good agreement was
obtained with the results of the MC simulations as presented in
Fig.~\ref{fig:effi-exp-th}.

When electrons from the target pass through the set-up to the wire chambers
multiple scattering in the target holder, in the wall of the carbon fiber 
vacuum chamber, and in the wire chamber windows takes place.  This gives rise to an
angular spread of the reconstructed angular correlation.

The simulated angular resolution corresponds to FWHM $\approx 7$ degrees. We
use bins of 10 degrees in the correlation spectra.

The shape of the coincidence response curve depends also on position of the
beam spot, which may walk during a long experiment. However, using the above
event mixing method, this effect can be compensated, so the extracted angular
correlation will be independent of small variations in the beam spot position.

\begin{figure}[htb]
\centering \includegraphics[width=70mm]{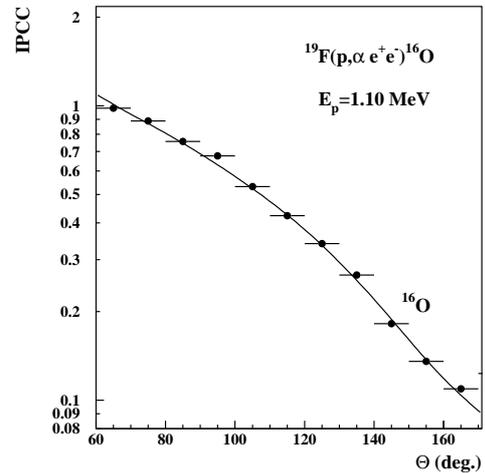}
\caption{\label{fig:e0-abra}Angular correlation of the e$^+$ e$^-$ pairs that
  originated from the $^{16}$O 6.05 MeV $E0$ transition excited in the
  $^{19}$F(p,$\alpha$)$^{16}$O reaction at E$_p$=1.10 MeV compared with the MC
  simulation assuming pure $E0$ transition.}
\end{figure}

In order to check the experimentally determined response curve with data, the
angular correlation of the e$^+$e$^-$ pairs created in the 6.05 MeV $E0$
transition was measured and corrected by the response curve determined in the
same experiment. As shown in Fig.~\ref{fig:e0-abra} very good agreement has
been obtained with the theoretically predicted $E0$ angular correlation.

\subsection{Background by cosmic muons}

Cosmic muons going through the spectrometer produce coincidences between the
$\Delta E-E$ telescopes and the MWPC detectors, similarly to e$^+$e$^-$
pairs. We measure low coincidence rates, especially at large separation
angles, so the effect of traversing cosmic rays has to be
considered. Background measurements have been performed before and after the
experiments with the settings (gates, thresholds, etc.) of the in-beam
measurement, and the angular correlation of the background events were
subtracted with a weighting factor. This factor was determined by comparing
the high energy part ($E_{sum}>20$ MeV) of the sum energy spectra measured
in-beam and off-beam, which contained only cosmic events in both cases.

\section{Measured pure $E1$ and $M1$ transitions}

To demonstrate the reliability of the spectrometer, we investigated a pure
$E1$ transition in $^{12}$C and and a pure $M1$ transition in $^{8}$Be as
well. The $^{12}$C resonance at 17.2 MeV with a width $\Gamma$=1.15 MeV is
populated in the $^{11}$B(p,$\gamma$)$^{12}$C reaction at 1.6 MeV bombarding
energy. It decays by isovector $E1$ transitions to the ground state and first
excited state with energies of 17.2 and 12.8 MeV. The $^8$Be resonance at 17.6
MeV with $\Gamma$=11 keV is populated in the $^7$Li(p,$\gamma$)$^8$Be reaction
at 441 keV proton bombarding energy. It decays to the ground state and the
particle-unstable first excited state ($\Gamma$=1.5 MeV) with 17.6 and 14.6
MeV isovector $M1$ transitions.

\begin{figure}[htb]
\centering
\includegraphics[width=70mm]{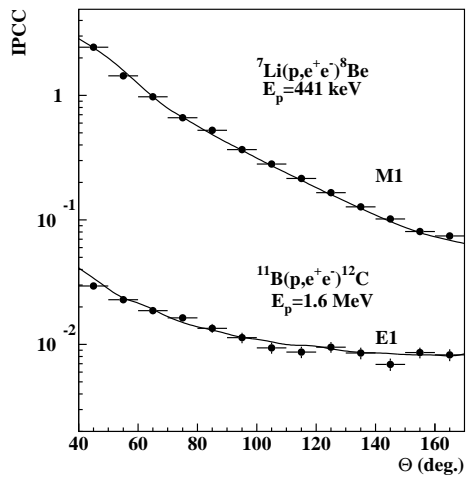}
\caption{\label{fig:e1-m1-abra}Measured and simulated angular correlations for
e$^+$e$^-$ pairs stem from a typical $E1$ ($^{12}$C 17.2 MeV) and $M1$
($^8$Be 17.6 MeV) transitions showing the very large discrimination power of
the IPCC for determining multipolarities. The values are rescaled for better
comparison of the shapes.}
\end{figure}

Figure \ref{fig:e1-m1-abra} shows the angular correlations for the above $M1$
and $E1$ transitions compared with the simulated full curves which confirms
the reliability of our setup. We have not observed significant anomaly for the
17.6 MeV isovector $M1$ transition in $^8$Be reported earlier \cite{bo96}.

\section{Outlook}

We repeated the experiment in order to investigate another $M1$ transition in
$^8$Be, which is isoscalar in contrast to the 17.6 MeV transition. The $^8$Be
resonance at 18.1 MeV ($\Gamma$=168 keV) was populated in the
$^7$Li(p,$\gamma$)$^8$Be reaction at 1.030 MeV proton bombarding energy. It
decays to the ground state and the particle unstable first excited state with
18.1 and 15.1 MeV isoscalar $M1$ transitions.

We have observed significant (5$\sigma$) deviation at $\Theta\approx135^\circ$
from the simulated angular correlation in the case of the 18.1 MeV isoscalar
$M1$ transition in $^8$Be with branching ratio relative to $\gamma$ ray
emission of $\approx5\times10^{-6}$. It can not be explained by any $E1$
admixture coming from the direct capture process. It has disappeared below and
above the 18.1 MeV resonance. However, it can be explained by the creation and
decay of a light ($m_0c^2$= 16.7 MeV) isoscalar J$^\pi$ = 1$^+$ boson
\cite{kr13}.

\section{Acknowledgements}

We are deeply indebted to Fokke W. N. de Boer, who proposed to search for a
short-lieved neutral boson in ATOMKI already in 2000. Together with him we
performed many challenging experiment in Debrecen. Fokke sadly passed away in
2010. This paper is dedicated to his memory. We are indebted also to Kurt
Stiebing for making his spectrometer available for our early experiments.
This work has been supported by the Hungarian OTKA Foundation No.\, K106035,
by the European Community FP7 - Capacities, contract ENSAR n$^\circ$ 262010
and by the European Union and the State of Hungary, co-financed by the
European Social Fund in the framework of T\'AMOP-4.2.4.A/2-11/1-2012-0001
‘National Excellence Program’.

%\end{linenumbers}
\end{document}